# Magnetic textures in a hexaferrite thin film and their response to magnetic fields revealed by phase microscopy

Atsuhiro Kotani[1], Ken Harada[2], Marek Malac[3,4], Hiroshi Nakajima[1], Kosuke Kurushima[5] and Shigeo Mori[1]

[1]*Department of Materials Science, Osaka Prefecture University, Sakai, Osaka 599-8531, Japan*

[2]*Center for Emergent Matter Science, The Institute of Physical and Chemical Research (RIKEN), Hatoyama, Saitama 350-0395, Japan*

[3]*Nanotechnology Research Centre, National Research Council (NRC), Edmonton, Alberta T6G 2M9, Canada*

[4]*Department of Physics, University of Alberta, Edmonton, Alberta T6G 2E1, Canada*

[5]*Toray Research Center, Otsu, Shiga 520-8567, Japan*

We investigated magnetic textures in a Sc-doped hexaferrite film by means of phase microscopy (PM) with a hole-free phase plate in a transmission electron microscope. In a zero magnetic field, the stripe-shaped magnetic domains coexist with magnetic bubbles. The magnetization in both magnetic domains was oriented perpendicular to the film and the domain walls have an in-plane magnetization. In the remnant state at 9.2 mT, several magnetic bubbles were formed with the formation of stripe-shaped magnetic domains, and the out-of-plane component in the stripe-shaped domains gradually appeared as the film thickness increased. As the film thickness increases further, the magnetic bubbles with clockwise or counter-clockwise spin helicities formed a triangular lattice. These results in the remnant state suggest that the domain wall energy in the magnetic bubble domains is lower in the thicker region.

# 1. Introduction

Phase microscopy (PM), especially with a hole-free phase plate (HFPP) in a transmission electron microscope (TEM), has been used to enhance phase contrasts of materials consisting of the light elements in biological fields [1–3]. PM has a potential advantage that the highly magnified images can be obtained *in-focus,* thus not suffering from Fresnel fringes caused by defocusing [4,5]. Unlike in electron holography, PM observation does not require a reference wave. Therefore, PM has been utilized for imaging the magnetization distribution of magnetic textures [6]. For example, skyrmions [7–10], vortex-like magnetic textures were observed using PM with an HFPP, and the semi-quantitative magnetic distribution was reproduced from the acquired PM image [11].

In this paper, we report that PM with an HFPP can be applied to observe the nanoscale magnetic textures, such as stripe-shaped magnetic domains and magnetic bubbles [12–16], in a uniaxial ferromagnet Sc-substituted *M*-type hexaferrite, $BaFe_{10.35}Sc_{1.6}Mg_{0.05}O_{19}$ (BFSMO). It has been revealed using Lorentz microscopy (LM) that magnetic bubbles, which have the vortex-like spin configuration similar to skyrmions, are formed by the application of the external magnetic field along the magnetic easy axis parallel to the *c* axis in BFSMO [17–19]. It can be, however, difficult to detect and interpret the spatial variations in the magnetization of samples using LM. The detailed magnetization distributions of magnetic domain structures are required to understand the formation mechanisms of magnetic bubbles. PM observations revealed semi-quantitative magnetization distributions of stripe-shaped magnetic domains, magnetic bubbles, and their responses to applied external magnetic fields. Moreover, the thickness dependence of the magnetization distribution in the magnetic textures in the remnant state was revealed in the PM observation.

# 2. Experimental methods

A single crystal of BFSMO specimens was synthesized via the floating zone method [20]. Specimens for TEM observation were thinned using Ar ion milling. The observations using PM with an HFPP were performed in a 300 kV TEM (Hitachi HF-3300). We utilized a 13 nm thick amorphous carbon film prepared by electron beam evaporation as an HFPP [1]. After the installation of the HFPP to the microscope column, the HFPP was heated to approximately 200 °C and kept at that temperature during the experiment to prevent the HFPP film from contamination [21,22].

We used the PM optics constructed in previous studies [11], as illustrated in Fig. 1. The HFPP was placed at the selected-area aperture plane, and the condenser lens was adjusted to construct the crossover at the HFPP. The application of magnetic fields perpendicular to the thin film was achieved by exciting the objective lens. The deviation of the crossover from the selected aperture position caused by exciting the objective lens was compensated by using the condenser lens. Image focusing was achieved by adjusting the excitation of the intermediate lens.

The phase shift $\Delta\phi$ is expressed as follows [23]:

$$\Delta\phi(x,y) = C_E V_0(x,y) t(x,y) - \frac{2\pi e}{h} \int \boldsymbol{B}(x,y) \mathrm{d}S. \quad (1)$$

Here, $C_E$ and $V_0$ and $t$, are the interaction constant which is 0.00652 rad V$^{-1}$ nm$^{-1}$ for 300 keV electrons, the mean inner potential, and the specimen thickness, respectively. $B$ is the magnetic flux of the specimen magnetization. The phase shift depends on both the mean inner potential and the specimen thickness. Assuming that the spatial change in the thickness can be negligible for the change in magnetization, the magnetization maps can be obtained from gradients of the phase distribution on the PM image acquired with an HFPP as follows [11]:

$$\boldsymbol{B} = \frac{h}{2\pi e}\Delta\phi(x,y) \propto \frac{h}{2\pi e}\Delta I(x,y) = \frac{h}{2\pi e}\left(\frac{\partial I_x}{\partial x}, \frac{\partial I_y}{\partial y}\right). \quad (2)$$

Here, $B$, $\phi$ and $I$, are the magnetic flux of the specimen magnetization, the phase shift due to the magnetic flux, and the intensity of the phase image. The absolute value of the in-plane magnetization can be obtained from the following equation,

$$|\boldsymbol{B}| = \frac{h}{2\pi e}\sqrt{\left(\frac{\partial I}{\partial x}\right)^2 + \left(\frac{\partial I}{\partial y}\right)^2}. \quad (3)$$

Note that the above equations are valid when an image recording device has a linear relationship between the number of detected electrons on the detector and the output intensity on the display, providing that the phase shift is proportional to the image contrast. Using these methods, we obtained magnetization maps from the PM images under magnetic fields applied externally.

## 3. Results and discussion

First, magnetic textures of BFSMO in zero magnetic field at room temperature were examined by PM observations. Figure 2(a) shows a PM image, showing the stripe-shaped magnetic domains and several magnetic bubbles. To extract the magnetization map of the BFSMO specimen in Fig. 2(b), differential images of the intensity in Fig. 2(a) were calculated according to Eq. (2). Figures 2(c) and (d) show the gradient images of the $x$ and $y$ directions, respectively. The white and black indicate positive and negative differential values, respectively. The magnetization vector map in Fig. 2(b) was obtained from Fig. 2(c) and (d). The color map indicates the direction of magnetization, coded according to the color wheel while the color saturation indicates the in-plane component of the magnetization intensity calculated using eq. (3).

In Fig. 2(b) the in-plane magnetization is indicated by white arrows. It shows the magnetization in the domain walls between the stripe-shaped magnetic domains is oriented parallel to the in-plane direction. In-plane component of the magnetization was not detected in the stripe-shaped magnetic domains and the magnetic bubbles, which suggests the magnetization in those domains is mainly oriented parallel to the easy axis (perpendicular to the thin film). Furthermore, it appears that the magnetization in the circular domain walls of the magnetic bubbles rotates clockwise (CW) or counterclockwise (CCW) in the plane of the thin film. The magnetization distribution of the magnetic bubbles obtained with PM is similar to that obtained by phase reconstructions through an iterative calculation using a series of 32 defocused images [24].

The magnetic texture varied depending on the strength of the applied external magnetic field. Figures 3(a) and (b) show the PM image and the magnetization map at 80 mT of the magnetic field applied, respectively. It can be seen in Fig. 3(a) that the width of the stripe-shaped magnetic domains is decreased and simultaneously the diameter of the magnetic bubbles is reduced as the strength of the magnetic field applied is increasing due to the Zeeman effect.

A Bloch line is formed, as indicated by the white arrowhead in Fig. 3(b). Note that a Bloch line is characteristic for the reversal of the domain-wall chirality, in which it has been accepted that the directions of the in-plane magnetization are reversed by gradual rather than abrupt rotation. The magnetization distribution obtained from Fig. 3(b) is schematically shown in Fig. 3(c). It has been recently recognized that the Bloch line plays an important role in the formation of magnetic bubbles [19]. As understood by comparing Figs. 2(b) and 3(b), a Bloch line was formed from the stripe-shaped magnetic domains by applying an external magnetic field.

As the strength of the magnetic field increases up to 120 mT, magnetic bubbles are formed from the stripe-shaped magnetic domains. Figure 3(d) shows the PM image of magnetic bubbles formed at 120 mT. Figure 3(e) is the magnetization map of magnetic bubbles indicated by the white dotted lines in Fig. 3(d). The magnetic bubbles have CW or CCW rotation of the magnetization. From the magnetization maps, the diameter of the magnetic bubble inside the circular domain wall can be estimated to be approximately 230 nm at 0 mT, 150 nm at 80 mT, and 100 nm at 120 mT, respectively. These results show that the diameters of magnetic bubbles decrease as the strength of the magnetic field increases.

After the external magnetic field up to 2 T was applied by exciting the objective lens, the lens was turned off to reduce the magnetic field quickly down to 9.2 mT. Figure 4 shows that in the remnant state numerous magnetic bubbles were formed in the region far from the edge of the specimen and the magnetic stripe domain exists in the region near the edge. The specimen in this study was made by ion milling and has a wedge-like thickness profile with progressively lower thickness closer to the edge. The thickness values in the region indicated by I, II, and III in Fig. 4 measured using electron energy-loss spectroscopy (EELS) are 39 nm, 96 nm, and 163 nm, respectively. Therefore, it is shown that the stripe-shaped magnetic domains are formed in the thinner region and magnetic bubbles are formed in the thicker region.

Here, the magnetization distribution of stripe-shaped magnetic domains and magnetic bubbles are discussed. PM observations revealed that changes in the magnetization distribution of the stripe-shaped magnetic domains and the transformation from the domains into magnetic bubbles with increasing the film thickness. Figure 5 shows the three magnetization maps of the regions I, II, III in Fig. 4 calculated using eq. (3). The white arrows indicate the magnetization direction. The magnetization map obtained from I shows that the magnetization in the stripe-shaped magnetic domain structure tends to the in-plane direction. Figure 4 shows that the out-of-plane component of magnetization in the magnetic domain increases as the film thickness increases. In the magnetization map II, the dark region where the magnetization is oriented perpendicular to the thin film in the stripe-shaped magnetic domain expands. These results show that the thickness dependence of the magnetization distributions of stripe-shaped magnetic domains can be clarified using PM with an HFPP.

For stripe-shaped magnetic domains, it has been reported that the thickness-driven reorientation of the magnetization from in-plane to perpendicular is caused [25,26] because the demagnetization field whose contribution favors an in-plane preferential orientation for the magnetization decreases as the thickness increases. The critical thickness, $t_C$, of the reorientation is given as follows:

$$t_C \sim 27.2\, M_S^2 / K_U^{3/2}, \qquad (4)$$

here $A$ is exchange stiffness constant, $M_S$ is saturation magnetization, and $K_U$ is uniaxial anisotropy constant. The parameters of the BFSMO specimen were reported to be $A = 1.3 \times 10^{-6}$ erg/cm, $M_S = 286$ emu/cm$^3$, $K_U = 5.3 \times 10^5$ erg/cm$^3$ [27], resulting in $t_C \sim 67$ nm. The PM observations in Fig. 5(I) and (II) experimentally show the reorientation of the magnetization across $t_C$.

As the film thickness increases to more than 100 nm, magnetic bubbles are formed. Figure 5(III) shows a magnetization map of the region III in Fig. 4. The magnetization distribution of magnetic bubbles with the CW or CCW spin rotation in the magnetic domain wall can be seen clearly, and those magnetic bubbles are formed locally in a triangular lattice. As shown in the magnetization maps in Figs. 2 and 3, black and white balls in Fig. 4 correspond to the magnetic bubbles with CW and CCW rotation of the magnetization, respectively. It can be seen in Fig. 4 that the same quantity of those magnetic bubbles exists and the magnetization helicities are oriented randomly. Therefore, it is suggested that the energies in the magnetic bubbles with the CW or CCW helicities are equivalent in the remnant state. It appears that the formation of magnetic bubbles in the remnant state depends on the film thickness.

It was reported that the magnetic energy of the stripe-shaped magnetic domain is higher than that of the magnetic bubble by magnetic domain-wall energy [28]. The wall energy $\sigma_W$ of a Bloch domain wall in a uniaxial ferrimagnet is expressed as follows [29]:

$$\sigma_W = 4\sqrt{AK}\ . \qquad (5)$$

The effective magnetic anisotropy $K$ related to $\sigma_W$ decrease as the thickness increases [30]. Thus, judging from eq. (4), the formation of magnetic bubbles in the remnant state is induced by lower $\sigma_W$ in the thicker region as shown in Fig. 5(III). The mechanism of thickness dependence in the formation of magnetic bubbles in the remnant state will be revealed in the future by the experiments with thickness-controlled films and theoretical simulations.

## 4. Conclusions

In conclusion, PM with an HFPP was utilized to reveal the magnetization distributions of magnetic textures, such as stripe-shaped magnetic domains and magnetic bubbles in BFSMO and their response to external magnetic fields applied perpendicular to the thin film. The stripe-shaped magnetic domains and a few magnetic bubbles coexisted in a zero magnetic field. PM observation revealed that the magnetic domain structures consist of the domain wall with the in-plane magnetization and the domain with the magnetization perpendicular to the film. In the remnant state after the external magnetic field was applied up to 2 T and was decreased, many magnetic bubbles with the stripe-shaped magnetic domains were observed. It was revealed the changes in magnetization distribution in the stripe-shaped magnetic domains as the film thickness increased. In addition, it was revealed that magnetic bubbles with the CW or CCW spin helicities were formed in a triangular lattice as the film thickness increased furthermore. The PM with an HFPP will be one of the powerful tools to analyze the magnetic distribution of complex magnetic textures.

## Acknowledgments

The authors greatly acknowledge Kai Cui and Mark Salomons for technical support during PM experiments and Jean Nassar for support of programming using Python codes. This work was partially supported by JSPS KAKENHI (No. 18J12180) and Graduate Program for System-inspired Leaders in Material Science (SiMS) of Osaka Prefecture University. This work was also supported by the National Research Council (NRC) in Canada. The experiments were made possible by outstanding support from Hitachi High Technologies for the Hitachi HF-3300 microscope at NRC-Nano.

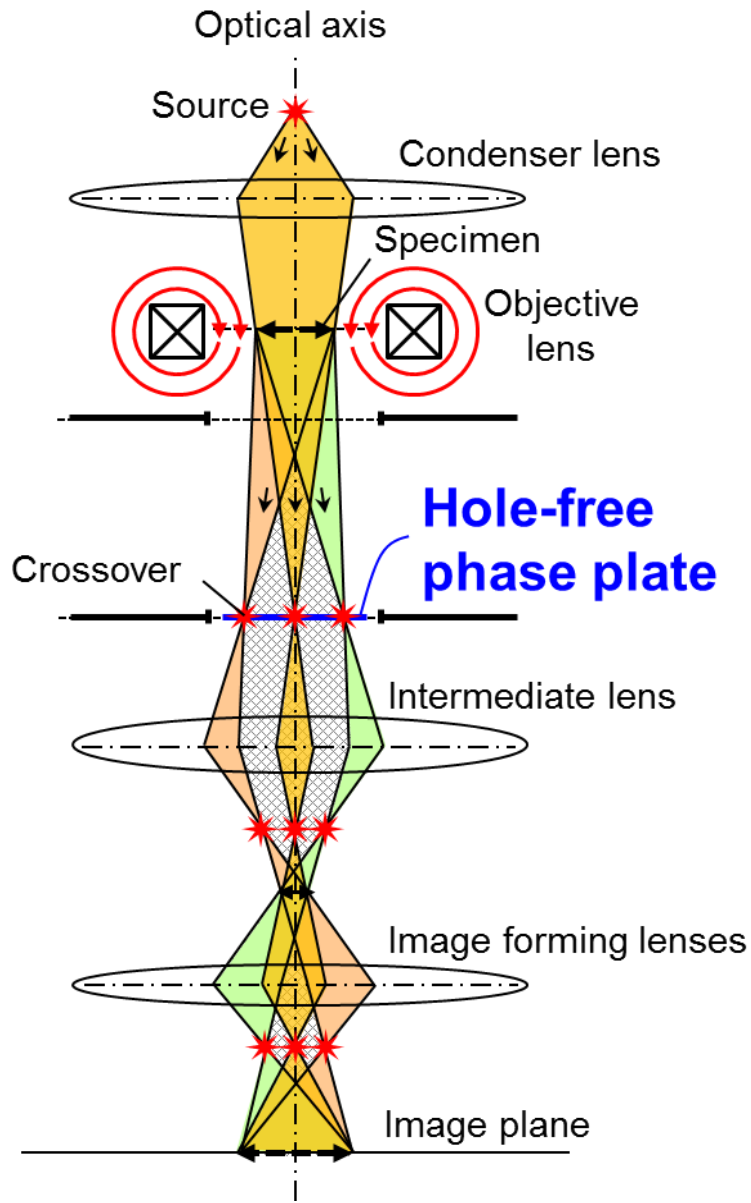

**Fig. 1.** A schematic illustration of the optic for PM using an HFPP. The red arrows indicate a magnetic field applied by a weakly excited objective lens.

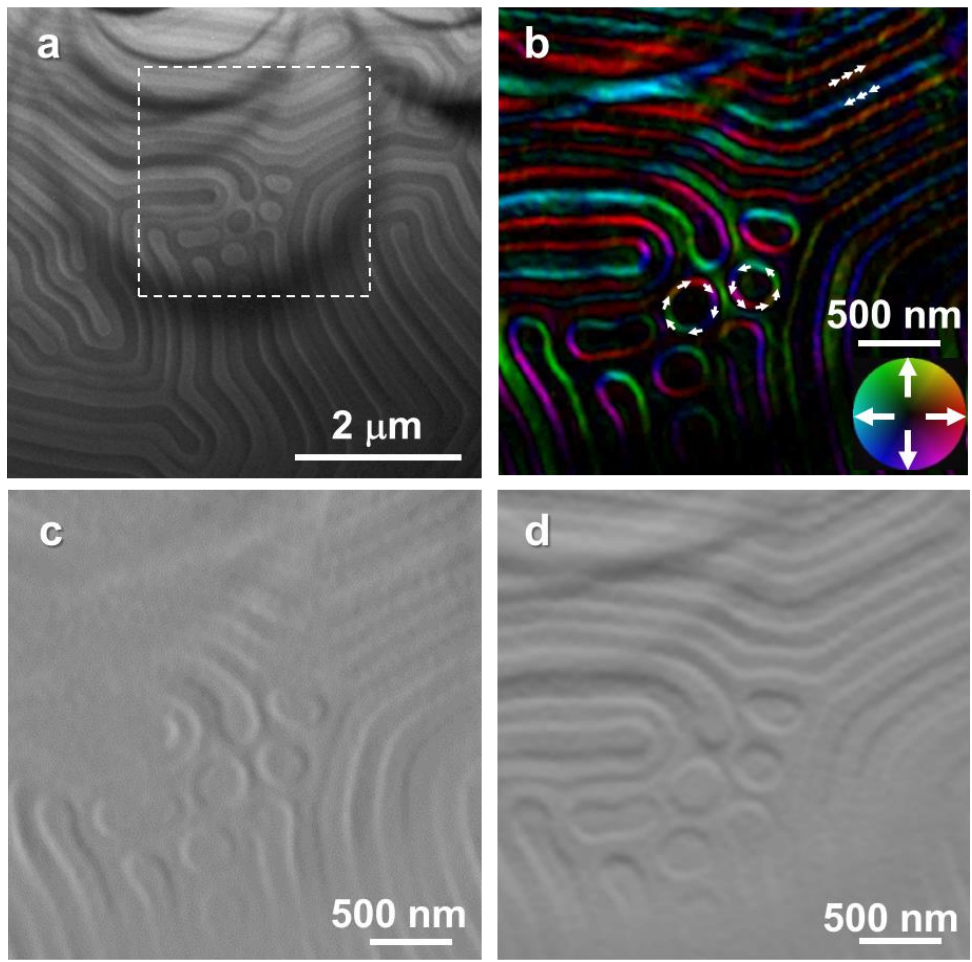

**Fig. 2**. (a) A PM image in a zero magnetic field at room temperature. (b) The magnetization map calculated from the differential images of (a) using the areas surrounded by the white dotted lines.   The color and the white arrows indicate magnetization direction as described by the color wheel. (c) and (d) showing the differential images in (a) in the *x* and *y* directions, respectively.

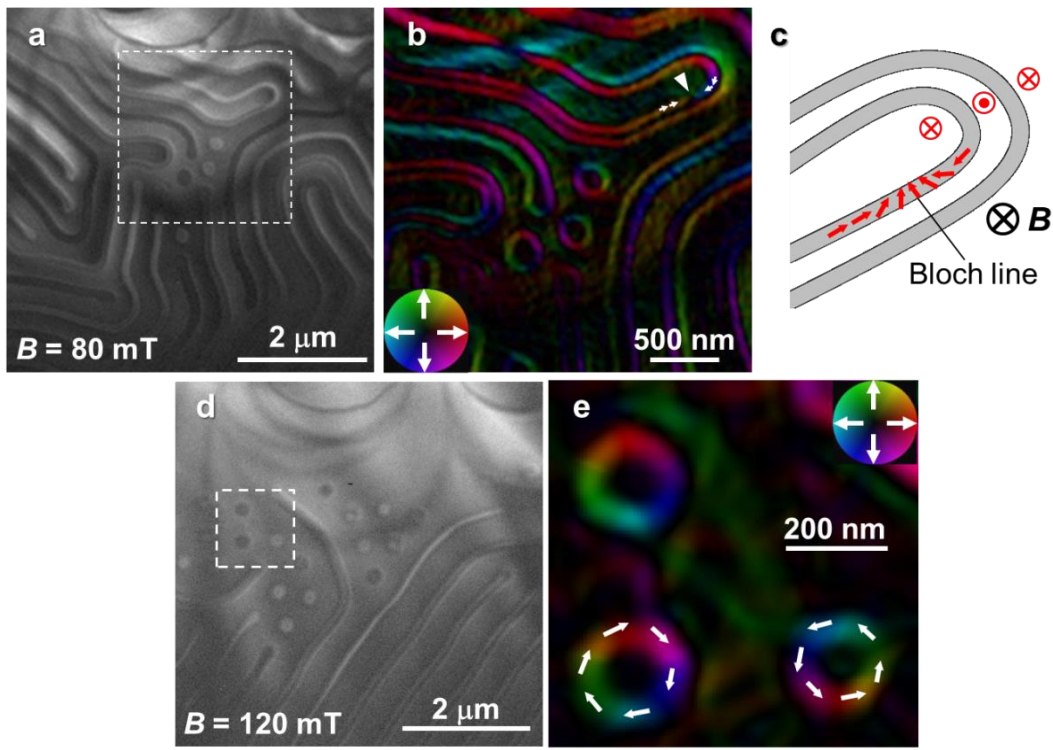

**Fig. 3.** (a) A PM image and (b) a magnetization map calculated by the differentiation of the intensity at 80mT using the region surrounded by the white dotted lines in (a). The region is the same as that in Fig. 2(b). The white arrows indicate the magnetization and the white arrowhead shows a Bloch line. (c) The schematic image of the magnetization distribution in (b), showing the Bloch line formed in the domain wall. The red arrows indicate the magnetic moments. (d) and (e) showing a PM image and a magnetization map of the area surrounded by the white dotted lines in (d) at 120mT.

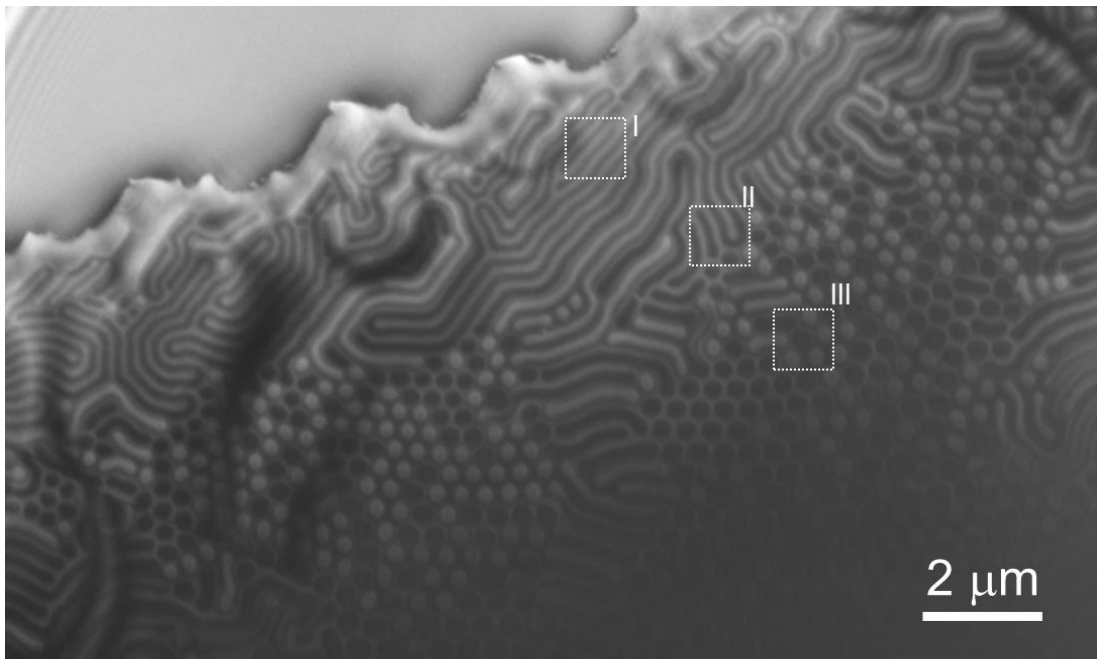

**Fig. 4.** The image of the magnetic bubbles formed in the remnant state after a rapid field change from 2T field saturation to 9.2 mT.

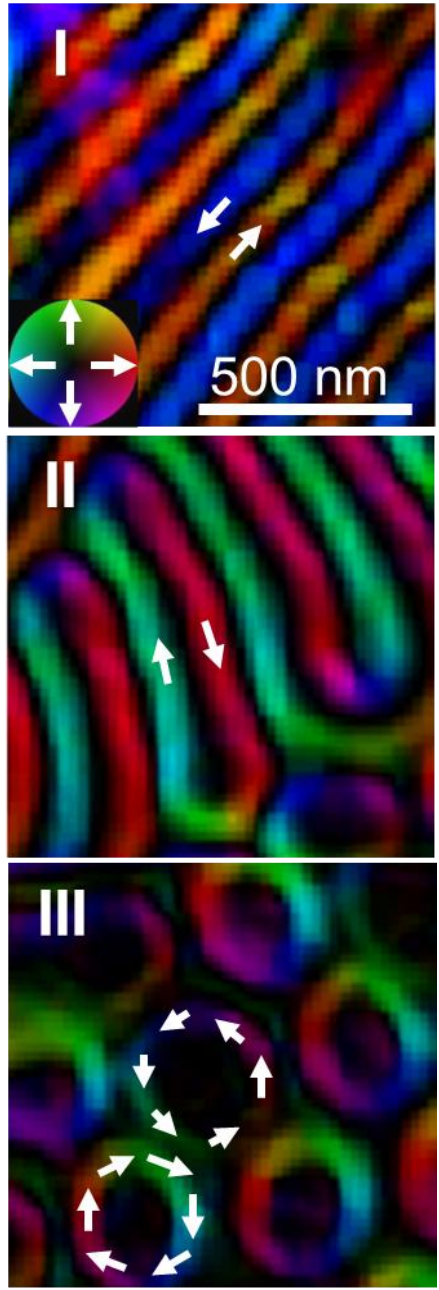

**Fig. 5.** The magnetization maps of the regions I, II, and III in Fig. 4.